\documentclass[useAMS,usenatbib,usedcolumn,usegraphicx]{mn2e}

\def\etal{{\it et\thinspace al.}\ }
\def\o4{{[{\sc O\,iv}]}}
\def\nev{{Ne\,{\sc v}}}
\def\o3{[{\sc O\,iii}]}

\def\lmlm{{$\lambda\lambda$}\ }
\def\mum{{$\mu$m}\ }

\def\eion{{(e~+~ion)}\ }
\def\te{{T$_e$}}
\def\ne{{N$_e$}}

\usepackage[dvips]{graphics}
\usepackage[dvipsnames,usenames]{color}

\newcommand{\be}{\begin{equation}}
\newcommand{\ee}{\end{equation}}
\newcommand{\cf}{{\rm cf.}\ }

\title{Fine structure collision
strengths and line ratios for [Ne\,{\sc v}] in infrared and optical sources}
\author
[Michael Dance, Ethan Palay, Sultana N. Nahar, Anil K.\ Pradhan]
       {Michael Dance\thanks{Present Address: Biomedical Program,
University of Toledo, Toledo, Ohio, USA.},Ethan
Palay\thanks{Corresponding author}, Sultana N. Nahar, Anil K.
Pradhan\\
       Department of Astronomy,
 The Ohio State University, Columbus, OH 43210, USA}
\date{Accepted  xxxxxx 
      Received xxxxxx;
      in original form xxxxxx}

\pagerange{\pageref{firstpage}--\pageref{lastpage}}
\pubyear{2012}

\def\LaTeX{L\kern-.36em\raise.3ex\hbox{a}\kern-.15em
    T\kern-.1667em\lower.7ex\hbox{E}\kern-.125emX}

\begin{document}

\maketitle

\label{firstpage}

\begin{abstract}
 Improved collisions strengths for the mid-infrared and optical
transitions in Ne\,{\sc v} are presented. Breit-Pauli R-Matrix calculations for
electron impact excitation are carried out with fully resolved
near-threshold resonances at very low
energies. In particular, the fine structure lines at 14 \mum and 24
\mum due to transitions among the ground state levels
$1s^22s^22p^3 \
^3P_{0,1,2}$, and the optical/near-UV lines at 2973, 3346 and 3426 $\AA$ 
transitions among the $^3P_{0,1,2}, ^1D_2, ^1S_0$ levels are described.
Maxwellian averaged collision strengths are tabulated for all
forbidden transistion within the ground configuration. Significant
differences are found in the low temperature range \te $<$ 10000 K for both
the FIR and the opitcal transitions compared to previous results.
An analysis of the
14/24 \mum ratio in low-energy-density (LED) plasma conditions 
reveals considerable
variation; the effective rate coefficient may be dominated by
the very low-energy behaviour rather than the maxwellian averaged
collision strengths. Computed values suggest a possible 
solution to the anomalous
mid-IR ratios found to be lower than theoretical
limits observed from planetary nebulae and Seyfert galaxies. 
While such LED conditions may be present in infrared sources,
they might be inconsistent with photoionization equilibrium models.  
   
\end{abstract}

\begin{keywords}
Gaseous Nebulae -- Optical Spectra: {\sc H\,ii} Regions -- Line Ratios: 
Atomic Processes -- Atomic Data
\end{keywords}

\section{Introduction}
 
 Several ionization stages of carbon-like ions provide useful
diagnostics in the mid- to far-infrared (MIR-FIR) space-borne
observations, as well as prominently observed optical lines from ground
based instruments. Nebular plasmas are generally the most common sources
of these lines arising from forbidden transitions within the levels of
the ground configuration of C-like ions (viz. Dopita and Sutherland 2003, 
Pradhan and Nahar 2011). Owing to much lower extinction than O/UV lines,
the MIR and FIR reveal spatially extended sources to much greater
depths. Therefore, in recent years space based observatories have
provided added impetus to the analysis of MIR and FIR lines observed
by {\it Spitzer}, {\it Herschel} and {\it Sofia}. In addition to
molecular clouds photoionized by hot stars into nebulae, a much wider variety
of astronomical objects whose physical conditions may be explored with
mid-IR [\nev] lines. 
These range from active
galactic nuclei (Melendez \etal 2011, Perez-Beaupuits \etal 2011, Stern \etal
2012, Smith \etal 2012), 
metal-rich {\sc H\,ii} regions (Furness \etal 2010), dusty
environments of Ultra-Luminous-Infrared-Galaxies (ULIRGs, Houck \etal
2004, 2005; Nagao \etal 2011), and blue compact dwarf galaxies (Izotov \etal
2012).

  H~II regions in general, and planetary nebulae (PNe) in
particular, are the testing ground as well as points of application of
forbidden line ratios diagnostics with respect to temperature, density,
and chemical abundances. The PNe are relatively bright and apparently 
simple objects that act as astrophysical laboratories to study basic
atomic processes. However, several previous studies have discussed
problems regarding [\nev] lines (Clegg \etal 1987, Oliva \etal 1996, van
Hoof \etal 2000). Based on the MIR 14/24 \mum line ratio, Rubin \etal
(2001) and Rubin (2004) pointed out that for a number of PNe the
observed values lie much lower than theoretically possible emissivity
ratios at standard nebular temperatures and densities.
More recently, in studies of spectral energy distributions in active
galactic nuclei (AGN) using high-ionization mid-IR lines, a number of 
observations also reveal anomalous 14/24
\mum line ratios found to be below the theoretical limit
(viz. Melendez \etal 2011, Tommasin \etal 2010, Weaver \etal 2010,
Pereira-Santaella \etal 2010). Although the line forming regions in
Seyfert galaxies may be dust obscured, the lower than predicted ratios are not
countenanced by derived mid-IR extinction (Li and Draine 2001).
It is important to address
this issue in order to use observed [\nev] line ratio as density diagnostics, as
the emissivity ratios indicate. Moreover, a deviation from theoretical
limits of the otherwise clean and uncluttered spectral [\nev] lines also
renders the determination of mid-IR extinction uncertain (Melendez \etal
2011).

 The analysis of the MIR and FIR space observations still rests on our
understanding of nebular astrophysics. Forbidden lines are the primary
diagnostics for low-temperature and low-density plasmas. In principle,
the atomic
physics is relatively straightforward: electron impact excitation of low-lying
levels followed by radiative decay via magnetic dipole or
electric-quadrupole transitions. These are the only two parameters
required for simple collisional-radiative models, usually involving no
more than the few levels of the ground configuration. However, it is
well known that the electron impact excitation collision strengths and
the relevant transition probabilities need to be calculated to high
accuracy in order to determine the temperature-density regimes in the
source. For example, C-like ground configuration $1s^22s^22p^3$ gives
rise to 3 LS terms and 5 fine structure levels $^3P_{0,1,2}, ^1D_2,
^1S_0$. In an earlier study, Palay \etal (2012)
showed that near-threshold resonances in
the collision strengths of C-like \o3 make significant contribution to fine
structure transitions. Rather elaborate computations including
relativistic and resonance effects are necessary in order to delineate
resonances and obtain accurate rate coefficients. In this work, we study
the transitions in the Ne\,{\sc v} ion and find similar effects.
As Rubin \etal (2001) and Rubin (2004) have noted, 
different sets of collision strength data available in literature (viz.
Lennon and Burke 1991, 1994; Griffin and Badnell 2000) did not
resolve the discrepancy in the 14/24 ratio lying below the theoretical
low density limit. The aim of this paper is to re-examine the
Ne\,{\sc v}  collision strengths and line ratios
computed using the Breit-Pauli-R-Matrix (BPRM) method (Berrington \etal
1995), with particular emphasis on
the MIR lines under low-energy-density (LED) conditions. In order to
examine the theoretical behaviour, we also extend the calculations to
to  \te $<$ 1000K, not considered in previous works.
Based on the more accurate data computed herein, the results
should enable revised interpretation of MIR [\nev] lines and originating
plasma environments.

\section{Theory and computations}
A brief theoretical description of the BPRM calculations is as follows
(a discussion of the general methodology and application to atomic
processes is given, for example, in Pradhan and Nahar 2011).

\subsection{The Breit-Pauli approximation} The relativistic BPRM
Hamiltonian  is given as

\begin{equation} 
\begin{array}{l}
H_{N+1}^{\rm BP} = \\ \sum_{i=1}\sp{N+1}\left\{-\nabla_i\sp 2 -
\frac{2Z}{r_i}
+ \sum_{j>i}\sp{N+1} \frac{2}{r_{ij}}\right\}+H_{N+1}^{\rm mass} + 
H_{N+1}^{\rm Dar} + H_{N+1}^{\rm so}.
\end{array}
\end{equation}
where the last three terms are relativistic corrections, respectively:
\begin{equation} 
\begin{array}{l}
{\rm the~mass~correction~term},~H^{\rm mass} = 
-{\alpha^2\over 4}\sum_i{p_i^4},\\
{\rm the~Darwin~term},~H^{\rm Dar} = {Z\alpha^2 \over
4}\sum_i{\nabla^2({1
\over r_i})}, \\
{\rm the~spin-orbit~interaction~term},~H^{\rm so}= Z\alpha^2 
\sum_i{1\over r_i^3} {\bf l_i.s_i}.
\end{array} 
\end{equation}

 Eq.~(2) representes the one-body terms of the Breit interaction. In
addition, another version of BPRM codes including the two-body terms
of the Breit-interaction (Pradhan and Nahar 2011; W. Eissner and G. X. Chen,
in preparation) has been developed, and is employed in the present work.

\subsection{Ne~V target representation}

 The coupled channel method embodied in the BPRM approximation is
crucially dependent on the accuracy of the wavefunctions of the
N-electron target
ion included to represent the (N+1)-electron \eion system. We employ the
general purpose atomic structure code SUPERSTRUCTURE (Eissner \etal
1974) to compute a  wavefunction expansion including the first 20
fine structure levels. Table~1 compares the calculated eigenenergies
with experimental values tabulated by the U.S. National Institute of
Standards and Technology (www.nist.gov). 

\begin{table}
\caption{Levels and energies ($E_t$) of target (core ion) Ne~V}
\begin{tabular}{rlrll}
\hline
\noalign{\smallskip}
& Level & $J_t$ & $E_t$(Ry) & $E_t$(Ry) \\
&  &  &NIST & SS \\
 \noalign{\smallskip}
 \hline
 \noalign{\smallskip}
1 & $1s^22s^22p^2(^3P)$   & 0   & 0.0     & 0. \\
2 & $1s^22s^22p^2(^3P)$   & 1   & 0.003758&0.0030391 \\
3 & $1s^22s^22p^2(^3P)$   & 2   & 0.010116&0.011366  \\
4 & $1s^22s^22p^2(^1D)$   & 2   & 0.276036 &0.30391   \\
5 & $1s^22s^22p^2(^1S)$   & 2   & 0.582424 &0.57413    \\
6 & $1s^22s2p^3(^5S^o)$   & 2   & 0.8052  &0.71604    \\
7 & $1s^22s2p^3(^3D^o)$   & 3   & 1.60232 &1.62957    \\
8 & $1s^22s2p^3(^3D^o)$   & 2   & 1.60296  &1.62932    \\
9 & $1s^22s2p^3(^3D^o)$   & 1   & 1.60316  &1.62929    \\
10& $1s^22s2p^3(^3P^o)$   & 2   & 1.89687  &1.92363     \\
11& $1s^22s2p^3(^3P^o)$   & 1   & 1.89687  &1.92340    \\
12& $1s^22s2p^3(^3P^o)$   & 0   & 1.89719  &1.92328    \\
13& $1s^22s2p^3(^1D^o)$   & 2   & 2.46556  &2.59326    \\
14& $1s^22s2p^3(^3S^o)$   & 1   & 2.54576  &2.64956    \\
15& $1s^22s2p^3(^1P^o)$   & 1   & 2.76854  &2.88988    \\
16& $1s^22s2p^4(^3P)$   & 2   & 3.76063  &3.86076    \\
17& $1s^22s2p^4(^3P)$   & 1   & 3.76778  &3.86807    \\
18& $1s^22s2p^4(^3P)$   & 0   & 3.77085  &3.87155    \\
19& $1s^22s2p^4(^1D)$   & 2   &          &4.13816    \\
20& $1s^22s2p^4(^1S)$   & 0   &          &4.74472    \\
            \noalign{\smallskip}
\hline
\end{tabular}
\end{table}

 The spectroscopic levels and configurations included in the (e~+~\nev)
wavefunction expansion are as in Table~1.  The full configuration-interaction
basis set optimized with SUPERSTRUCTURE is (Eissner \etal 1974, Nahar
\etal 2003): $[1s^2] 2s^22p^2, 2s2p^3, 2s^22p3s, 2p^4, 2s^22p3p,
2s^22p3d,\\
2s^22p4s, 2s^22p4p, 2s2p^23s,\\ 2s2p^23p, 2s2p^23d, 2s^23s^2,
2s^23p^2,
2s^23d^2, 2s^24s^2, 2s^24p^2,\\ 2s^23s3p, 2s^23s4s, 2s^23p3d, 2p^33s,
2p^33p, 2p^33d$. Although the computed enegies are generally within a 
few percent of the observed values, the latter are used in the 
BPRM calculations to ensure precise 
resonance positions relative to the Ne\,{\sc v} target thresholds.  
The collision strengths were computed employing the extended 
BPRM codes including a full representation of the two-body
Breit terms (Eissner and Chen 2012). 
It is important to ensure convergence of collision strengths with respect to
partial waves and energy resolution. Total \eion symmetries up to
(LS)J$\pi$ with J $\leq 19.5$ were included in the calculations, though
it was found that the collision strengths for all forbidden transition
transitions converged for J $\leq 9.5$. A very fine energy mesh of $\Delta E <
10^{-6}$ Rydbergs was used to resolve the near-thresold resonances.

\subsection{Effective collision strengths}

The effective collision
strengths or rate coefficients 
are computed by convolving the collision strengths over a Maxwellian
function at a given electron temperature \te as

\be 
\Upsilon_{ij}(T_e) = \int_0^{\infty} \Omega_{ij} (\epsilon)
\exp(-\epsilon/kT_e) d(\epsilon/kT_e), \ee

 where $E_{ij}$ is the energy difference and $\Omega_{ij}$ is the
collision strength for the transition $i \rightarrow j$. Later we
discuss LED plasma conditions where the effective collision strength may
differ from the form in Eq.~(3). The excitation rate coefficient is
related to the effective collision strength as

\be q_{ij}(T_e) = \frac{8.63 \times 10^{-6}}{g_i T_e^{1/2}}
e^{-E_{ij}/kT_e} \Upsilon_{ij} (T_e), \ee

where $g_i$ is the statistical weight of the initial level.

\begin{figure} 
\centering  
\includegraphics[width=\columnwidth,keepaspectratio]{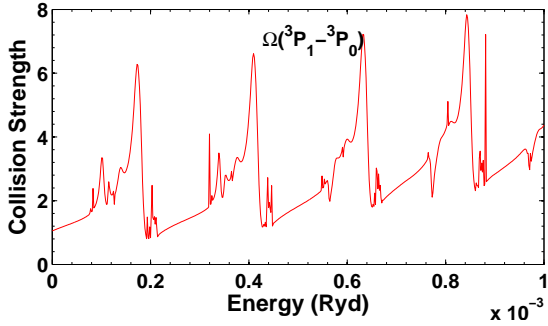}
\caption{Rydberg series of near-threshold resonances in the collision strength
for the fine structure transition $2p^2 (^3P_0-^3P_1)$
corresponding to the 24 \mum FIR line. 
A fine mesh order $10^{-6}$ Ryd was used to 
fully resolve the $(^3P_2, ^1D_2, ^1S_0) n \ell$ resonances, 
necessary for accurate rate coefficients at 
low temperatures. \label{fig:collryd}} 
\end{figure}

\begin{figure} 
\centering
\includegraphics[width=\columnwidth,keepaspectratio]{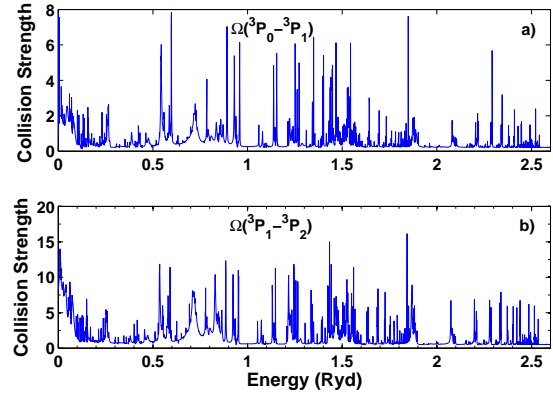}
\caption{Collision strengths over extended energy ranges
for fine structure transitions
$2p^2 (^3P_0-^3P_1,^3P_1-^3P_2)$ at  a) 24 \mum b) 14 \mum
respectively.
\label{fig:collfs}}
\end{figure}

\section {Results and discussion}

 Collision strengths and line ratios are presented in this section. A
discussion of possible implications for LED plasmas in general, and PNe
in particular, is also given.

\subsection{Ground state fine structure collision strengths}

 One of the main aims of the present calculations was complete
resolution of resonance structures in near-threshold collision
strengths. Fig.~\ref{fig:collryd} shows the resolved Rydberg series of
resonances in $\Omega(^3P_0-^3P_1)$ converging on to the 
higher $^3P_2, ^1D_2, ^1S_0$ levels. 
The calculations were carried out at an energy interval
of 10$^{-6}$ Ryd. These resonances dominate the 
low-energy-temperature behaviour of the detailed, as well as the averaged,
collision strengths. The corresponding 24 \mum line is the one most
affected. Fig.~\ref{fig:collfs} shows the results over an extended 
energy range for $\Omega(^3P_0-^3P_1)$ and $\Omega(^3P_1-^3P_2)$.
Particularly noteworthy is the significant rise close to thresholds for
both transitions.

\begin{figure} 
\centering
\includegraphics[width=\columnwidth,keepaspectratio]{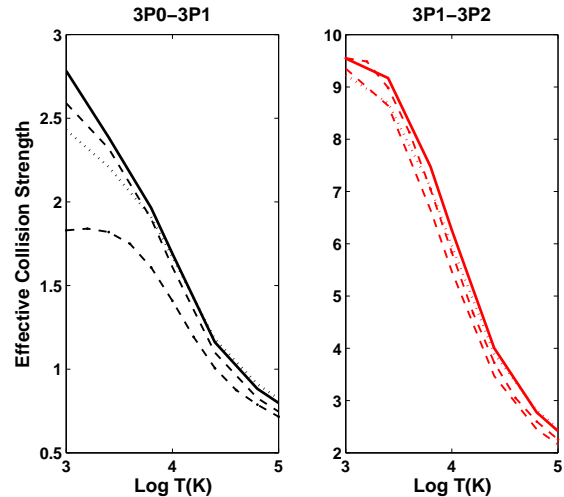}
\caption{Maxwellian averaged effective
collision strengths $\Upsilon(T_e)$ 
for the transitions $2p^2 (^3P_0-^3P_1, ^3P_1-^3P_2)$ at a)  24 \mum
 b) 14 \mum (\cf Fig.~2): solid line.
Dotted lines show previous results without
relativistic effects (Lennon and Burke 1994); dash-dot and dashed lines
show 20-level BPRM and 138-level ICFT results respectively (Griffin and Badnell
2000), available in the temperature range T$_e >$ 1000 K. 
The enhancements in present calculations
are due to the
fully resolved resonance structures near the threshold (\cf Fig.~1).
\label{fig:upsfs}}
\end{figure}

 The Maxwellian averaged collision strengths are plotted in
Fig.~\ref{fig:upsfs} (solid curve), 
and compared with previous results from Lennon and
Burke (1994, dotted curve) and two sets of calculations by Grfiffin and Badnell
(2000). The latter work employed two different approximations,
a 20-level BPRM calculation (dash-dot curve)
and a more extensive intermediate-coupling
fine structure transformation (ICFT, dashed curve) that is essentially
non-relativistic (as in Lennon and Burke 1994) 
but incorporates relativistic terms perturbatively via
an algebraic transformation. For the 24 \mum transition, we 
find differences of up to
30\% at the lowest temperatures with the otherwise similar previous 
20-level BPRM results, most likely attributable to improved
resolution at very low energies in the present calculations. 
Differences no larger than 10\% are
found for the 14 \mum transition between all sets of calculations.

 Present calculations are also carried out at \te $<$ 1000K, not
considered in previous works. Since we have taken particular care to
fully resolve the near-threshold collision strengths, the effective
collision strengths should be limited only by the intrinsic accuracy of
the BPRM calculations and not by numerical resolution of resonances.
That is necessary in order to interpret the line ratio observations in
cases where they do not fall in the nebular temperature range, as
discused next.

\subsection{The 14/24 \mum emissivity ratio and observations}

\begin{figure} 
\centering
\includegraphics[width=\columnwidth,keepaspectratio]{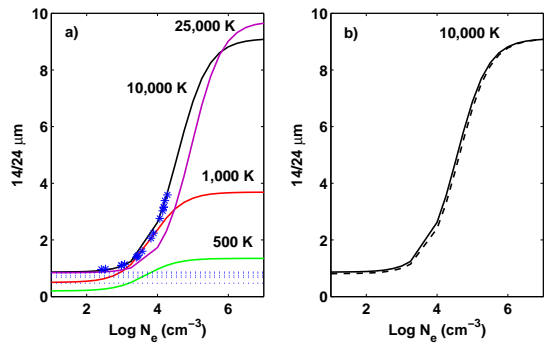}
\caption{The density and temperature dependence of the 14/24 \mum
line emissivity ratios: a)  Solid lines represent line ratios at four
temperatures. Asterisks denote
observed line ratios from planetary nebulae (Rubin 2004), 
at \te = 10,000 K with assigned
densities. Dotted lines represent
observed line ratios that are out of range of typical nebular
temperature-density range (Rubin 2004), except at much lower
temperatures. b) solid curves
are present results and the dashed lines are using the Lennon and
Burke  (1994) collision strengths.  Large differences in line ratios
are not found because of {\it systematic} differences of similar
magnitude for both lines between previous and present rate coefficients.
\label{fig:14/24}} 
\end{figure}

 Owing to small differences in excitation energies, the fine structure
line ratios due to transitions within the ground state levels 
$^3P_{0,1,2}$ are not highly temperature sensitive in the typical
nebular range around 10,000K. On the other hand, this emssivity ratio is highly
density sensitive at typical nebular densities of 10$^{3-6}$ cm$^{-3}$. 
 Rubin (2004) has tabulated the 14/24 line flux ratio from a number of
PNe, and noted that it falls {\it below} the theoretically possible
value of approximately 1.0 in about half the observations. 
Fig.~\ref{fig:14/24} shows the measured line ratios; the asterisks
denote the derived densities for the PNe that fall very well on the
10,000 K emissity ratio curve. However, the observed values from some of 
the anomalous PNe are seen as lying around the emssivity ratio
curve at 1000K or lower, shown as vertical dashed band in
Fig.~\ref{fig:14/24}. One can readly find low 
temperature curves vs. density that could account for these ratios lying
below the 10,000K curve.  
Incidentally, variations by similar amounts in both
transitions yield very good agreement with ratios at \te $>$ 10,000K
with previous works (viz. Lennon and Burke 1994), shown in the right
panel.

 Based on this work we can rule out further uncertainties in the collision
strengths. Fig.~\ref{fig:14/24} illustrates the 
dichotomy between the PNe with normal density sensitivity of the 14/24  
ratio, and a fair number of those that appear to correspond to much
lower temperatures than that which characterizes typical nebular
photoionization equilibrium. While the equilibrium models may not
explain these values, it is worth examining the behaviour of line ratios
in LED plasmas in general. Eq.~(3) implies that

\be lim_{T \rightarrow 0} \Upsilon(T) = lim_{E \rightarrow 0} \Omega(E). \ee 

 Between 1000-10,000K the Mawellian factor involves exponential
factor exp(-E/k\te) ranging from  exp(-158 E) to exp(-15.8E) in Rydberg
units. Therefore $\Upsilon(T)$ depends entirely on the detailed
behaviour of $\Omega(E)$. Furthermore, if we also consider that fact
that low densities \ne $\approx 10^2$ cm$^{-3}$ may exist, a departure
from the normal Maxwellian distribution under LED conditions is
possible. Therefore, the rate coefficient would tend to follow the
collision strengths {\it and} variations with resonance structures at
very low energies close to exciation threshold(s). This suggestion is
offered to re-examine non-equilibrium conditions that might prevail in
LED plasma sources, which may also not have 
established a Maxwellian distribution. The
emitted line intensity would then depend on the actual kinetic energy 
and number density of electrons at a given energy. In addition, the
electron and ion temperatures may differ, i.e. \te $\neq$ T$_i$. 

\subsection{[\nev] optical collision strengths and line ratios}
\begin{figure} 
\centering
\includegraphics[width=\columnwidth,keepaspectratio]{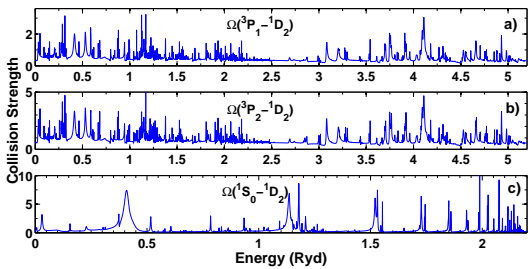}
\caption{Collision strengths for the forbidden optical transitions $2p^2
(^3P_1-^1D_2, ^3P_2-^1D_2, ^1S_0-^1D_2)$ at 3346,3426,2973 $\AA$
respectively \label{fig:collopt}}.
\end{figure}

\begin{figure} 
\centering
\includegraphics[width=\columnwidth,keepaspectratio]{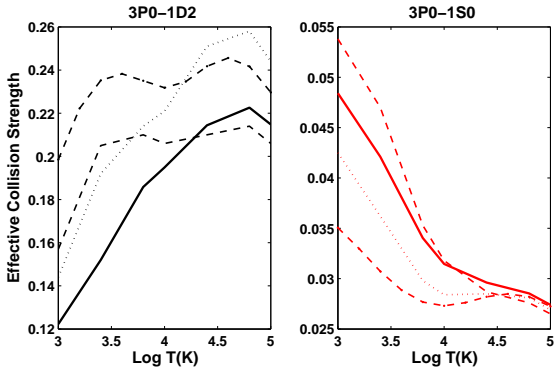}
\caption{Maxwellian averaged collision strengths $\Upsilon(T_e)$ for
the transitions $2p^2 (^3P_0-^1D_2, ^3P_0-^1S_0)$: Solid lines.
Dotted lines show previous results without relativistic effects (Lennon
and Burke
1994); dashed and dash-dot lines show 20-level BPRM and 138-level ICFT
results respectively (Griffin and Badnell 2000)
\label{fig:upsopt}}

\end{figure}

\begin{figure} 
\centering
\includegraphics[width=\columnwidth,keepaspectratio]{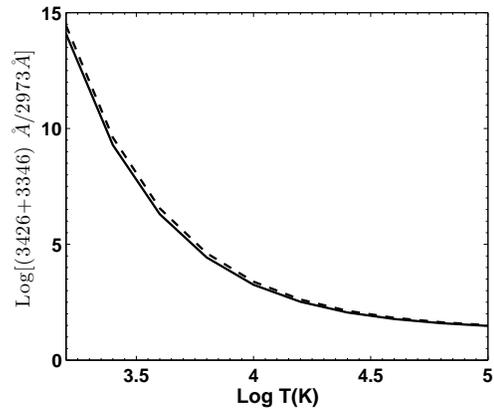}
\caption{Blended [\nev] optical line ratio (3346+3426)/2973 vs. \te at
\ne = 10$^3$ cm$^{-3}$. Despite significant differences in the collision
strengths the optical emissivity ratios agree with those using
earlier data by Lennon
and Burke (1994) owing to systematic enhancements in the same direction.
\label{fig:lrcomp}}

\end{figure}

 Collision strengths for forbidden optical transitions within ground
configuration levels are shown in Fig.~\ref{fig:collopt}:
$^3P_1-^1D_2$, $^3P_2-^1D_2,^1S_0-^1D_2$ at \lmlm 3345.8, 3425.9, 2972.8, 
respectively. Resonance structures due to Ne\,{\sc v} 
target excitation thresholds
given in Table~1 manifest themselves at all energies, and need to be
considered in the calculation of averaged collision strengths. Whereas
the agreement with previous works (Lennon and Burke 1994, Griffin and
Badnell 2000) is found to be satisfactory, generally 10-15\% or better, 
for some of the relatively weak transitions there are marked 
differences. Fig.~\ref{fig:upsopt} compares the present results for the
maxwellian averaged collision strengths for two transitions,
$^3P_0-^1D_2$ and $^3P_0-^1S_0$ that do not directly correspond to
observed lines but enter in the solution of rate equations for line
emissivities. The differences are found to be rather large.
Nevertheless, the important optical line ratios are not significantly
affected, as shown in Fig.~\ref{fig:lrcomp}. The blended emissivity
ratio (3426+3346)/2973 differs little from those derived from earlier
values from Lennon and Burke (1994).

\subsection{Maxwellian averaged collision strengths}

 In Table~1, we present the effective collision strengths (Eq. 3) for the 10
transitions among the ground configuration levels and their wavelengths.  
The tabulation is carried out at a range of temperatures typical of
nebular environments, including the low temperature range $T \leq 1000$K 
not heretofore considered. 

\begin{table*}
\begin{minipage}{148mm}
\caption{Effective Maxwellian averaged collision strengths}
\begin{tabular}{ccccccccc}
\hline
Transition & $\lambda$ & $\Upsilon$(100) & $\Upsilon$(500) &
$\Upsilon$(1000) & $\Upsilon$(5000) & $\Upsilon$(10000) & 
$\Upsilon$(20000) & $\Upsilon$(30000)\\
\hline
 $^3P_0-^3P_1$ & 14 $\mu$m & 2.72(0) & 3.02(0) & 2.78(0) & 2.09(0)
& 1.69(0) & 1.28(0) & 1.07(0)\\
 $^3P_2-^3P_0$ & 9 $\mu$m & 2.70(0) & 3.00(0) & 3.16(0) &
2.56(0) & 1.93(0) & 1.34(0) & 1.04(0)\\
 $^3P_2-^3P_1$ & 14 $\mu$m & 8.48(0) & 9.18(0) & 9.55(0) &
8.02(0) & 6.26(0) & 4.49(0) & 3.60(0)\\
 $^1D_2-^3P_0$ & 3301.3 $\AA$ & 1.23(-1) & 1.20(-1) & 1.22(-1) &
1.80(-1) & 1.95(-1) & 2.09(-1) & 2.19(-1)\\ 
 $^1D_2-^3P_1$ & 3345.8 $\AA$ & 3.76(-1) & 3.67(-1) & 3.69(-1) &
5.37(-1) & 5.84(-1) & 6.27(-1) & 6.58(-1)\\
 $^1D_2-^3P_2$ & 3425.9 $\AA$ & 6.61(-1) &6.46(-1) & 6.47(-1) &
9.22(-1) & 9.97(-1) &1.07(0) & 1.1(0)\\
 $^1S_0-^3P_0$ & -  & 4.60(-2) & 4.82(-2) & 4.84(-2) &
3.58(-2) & 3.15(-2) & 2.98(-2) & 2.94(-2)\\
 $^1S_0-^3P_1$ & 1574.8 $\AA$ & 1.35(-1) & 1.42(-1) & 1.42(-1) &
1.04(-1) & 9.13(-2) & 8.73(-2) & 8.69(-2)\\
 $^1S_0-^3P_2$ & 1592.3 $\AA$ & 2.24(-1) & 2.35(-1) & 2.35(-1) &
1.73(-1) & 1.53(-1) & 1.47(-1) & 1.45(-1)\\
 $^1S_0-^1D_2$ & 2972.8 $\AA$ & 3.51(-1) & 4.51(-1) & 4.88(-1) &
6.69(-1) & 6.26(-1) & 6.33(-1) & 6.87(-1)\\
\hline
\end{tabular}
\end{minipage}
\end{table*}

\subsection{Conclusion}

 New calculations for [\nev] collision strengths are reported including
relativistic and resonance effects at very low energies that affect the
low temperature rate coefficients not heretofore calculated.
In particular, a precise delineation of
near-threshold resonance structures has a significant effect on the
mid-IR 14/24 \mum line emissitivies. However, low-density theoretical limits 
of the ratios computed at typical
nebular temperatures \te $\approx$ 10,000K are still higher than the 
anomalously low values of observations
from a number of PNe discussed by Rubin \etal (2001) and
Rubin (2004). Since the Maxwellian averaged collision strengths
generally decrease with temperature, and simulate the behaviour of
collision strengths at low energies, it is suggested that
non-equilibrium conditions corresponding to LED plasma conditions might
possibly account for the low observed 14/24 ratios. In any case, the low
temperature rate coefficients for \te $<$ 10,000K are likely to be
uncertain owing to resonances at low energies.

 The [\nev] forbidden optical collision strengths are generally in good
agreement with previous works, and the prominent line ratios are not
much influenced.
Finally, as noted in the previous work on \o3 lines (Palay \etal 2012),
for higher
temperatures T $>$ 20,000K proton impact excitation of the ground state
fine structure levels $^3P_{0,1,2}$ needs to be taken into account;
at lower temperatures the excitation rate coefficent due
to electrons far exceeds that due to protons (Ryans \etal 1999).

 Collision strengths for all 190 transitions among energy levels in
Table~1 may be obtained from Ethan Palay (palay.5@buckeyemail.osu.edu).

\section*{Acknowledgments}
 The computational work was 
carried out at the Ohio Supercomputer Center in Columbus 
Ohio. This work was partially supported by a grant from the NASA
Astrophysical Research and Analysis program. EP would like to
gratefully acknowledge a Summer Undergradute Research Program grant from the
Ohio State University.

\label{lastpage}

\end{document}